\documentclass[aps,prl,twocolumn,showpacs,amsmath,amssymb]{revtex4}

\usepackage{graphicx}
\usepackage{color}

\begin{document}

\title{Non-Local Product Rules for Percolation}

\author{Saulo D. S. Reis}
\email{saulo@fisica.ufc.br}
\author{Andr\'e A. Moreira}
\email{auto@fisica.ufc.br}
\author{Jos\'e S. Andrade Jr.}
\email{soares@fisica.ufc.br}
\affiliation{Departamento de F\'{\i}sica, Universidade Federal
do Cear\'a, 60451-970 Fortaleza, Cear\'a, Brazil}

\date{\today}

\begin{abstract}
  Despite original claims of a first--order transition in the product
  rule model proposed by Achlioptas {\it et al.} [{\it Science} {\bf
    323}, 1453 (2009)], recent studies indicate that this percolation
  model, in fact, displays a continuous transition. The distinctive
  scaling properties of the model at criticality, however, strongly
  suggest that it should belong to a different universality class than
  ordinary percolation. Here we introduce a generalization of the
  product rule that reveals the effect of non--locality on the
  critical behavior of the percolation process. Precisely, pairs of
  unoccupied bonds are chosen according to a probability that decays
  as a power-law of their Manhattan distance, and only that bond
  connecting clusters whose product of their sizes is the smallest,
  becomes occupied. Interestingly, our results for two-dimensional
  lattices at criticality shows that the power-law exponent of the
  product rule has a significant influence on the finite-size scaling
  exponents for the spanning cluster, the conducting backbone, and the
  cutting bonds of the system. In all three cases, we observe a
  continuous variation from ordinary to (non-local) explosive
  percolation exponents.
\end{abstract}

\maketitle
The percolation paradigm represents a formidable example where a
simple geometrical construction leads to profound concepts in
statistical physics, with special emphasis on phase transitions, and
real applications in science and technology
\cite{Stauffer1992,Sahimi1994,Perclass}. Standard percolation
processes are based on local rules, since they are accomplished
through random allocation of sites or bonds, therefore disregarding
any spatial correlation or global information involved in the
occupation of other elements on the lattice. However, in the case of
long-range spatially-correlated percolation
\cite{Percorr,Prakash1992}, the probability for a site to be occupied
depends on the occupancy of other sites. Moreover, it has been shown
that spatial long-range correlations in site occupancy can give rise
to important changes on the structural characteristics of the spanning
cluster as well as its corresponding conducting backbone
\cite{Prakash1992}.  These changes are strong enough to modify the
scaling exponents of traditional (local) percolation.

Recently, a new percolation model has been proposed, the so-called
Product Rule (PR) percolation, in terms of a bond occupation process
that is essentially non-local \cite{Achlioptas2009}. In this model, at
each step, two unoccupied bonds are randomly chosen and associated
with weights given by the product of the cluster sizes they would
potentially connect. Only that bond which has the smallest weight
becomes occupied. By comparison with the traditional percolation model
\cite{Stauffer1992}, the PR model presents a more abrupt transition
when applied to different network topologies
\cite{Percexp,Friedman2009,Manna2011,Radicchi2009,Radicchi2010,Ziff2009,Ziff2010,Andrade2011}.
As potential applications, the PR model has been recently associated
to the growth dynamics of Protein Homology Networks
\cite{Rozenfeld2010} as well as to the formation of bundles of
single-walled nanotubes \cite{Kim2010}.

\begin{figure}[!t]
  \includegraphics[width=6cm]{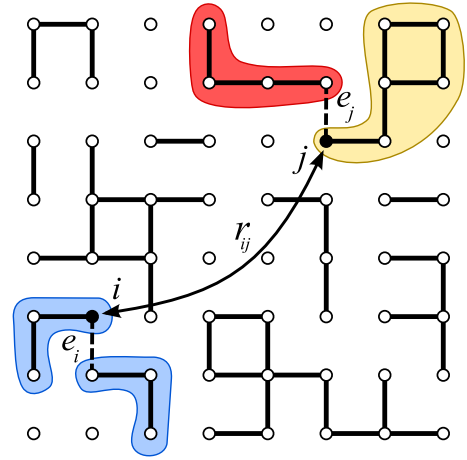}
  \caption{(Color online) Pair of unoccupied bonds $e_{i}$ and $e_{j}$
    (dashed lines) randomly selected for the application of the
    product rule, according to the probability $P(r_{i})\sim
    r_{j}^{-\alpha}$, where $r_{ij}$ is the Manhattan distance between
    sites $i$ and $j$ (black circles), and $\alpha$ is a variable
    exponent. Following the PR model, the bond $e_{i}$, merging the
    two clusters in blue (with 3 sites each), becomes occupied. The
    bond $e_{j}$ would merge the the clusters in red ($4$ sites) and
    yellow ($6$ sites), but remains unoccupied.}
  \label{fig1}
\end{figure}

Regardless of initial claims of a first order transition in the PR
model \cite{Achlioptas2009}, however, recent analytical and numerical
works \cite{Grassberger2011,Percont} have demonstrated that the
alleged ``Explosive Percolation'' process actually displays a
continuous, i.e., a second-order phase transition.  This apparent
drawback of the PR model has been somehow overstated, in the sense
that the model proposed by Achlioptas {\it et
  al.}~\cite{Achlioptas2009} certainly represents an original and
interesting contribution to the field. For instance, much less
importance has been given to the non-local attributes of the PR
algorithm. As a consequence of this non-locality, the model exhibits a
percolation transition that, although continuous in nature, seems to
belong to a different universality class than ordinary percolation
\cite{Ziff2009,Ziff2010,Andrade2011}. Under this framework, the
adjacent edge rule (AER) model~\cite{DSouza2010} represents a
particular case of the PR process that is analytically tractable,
since the selection is restricted to two adjacent bonds. When applied
to random graphs, this model still displays a more abrupt transition
and different scaling properties than ordinary percolation.

\begin{figure}[!t]
  \includegraphics[width=8cm]{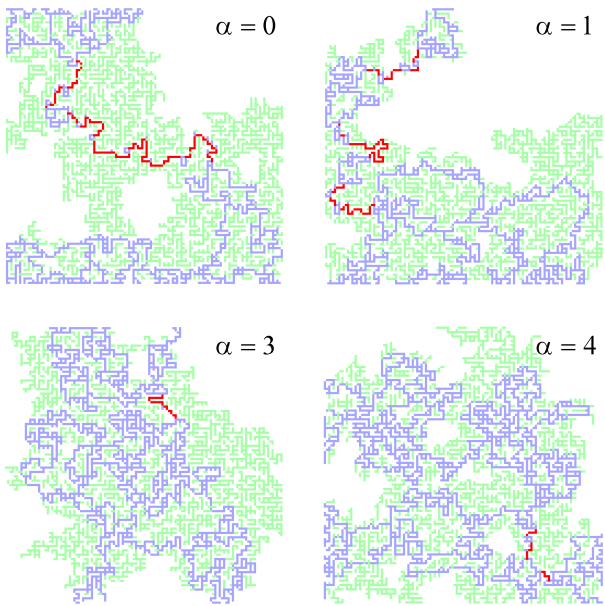}
  \caption{(Color online) Snapshots of the largest cluster at $p_{c}$
    for different values of the exponent $\alpha$, and a lattice size
    $L=64$. The bonds forming the conducting backbone are
    in blue, the cutting bones are in red, and the
    remaining bonds of the largest cluster are presented in green.
    Although no major difference can be observed on the mass
    $M_{clus}$ of the largest cluster, one can notice that the
    conducting backbone occupies a larger fraction of the largest
    cluster as $\alpha$ increases, leading to a substantial decrease
    on the number of cutting bones $M_{cut}$.}
  \label{fig2}
\end{figure}

In this Letter we introduce a generalization of the PR model in which
the range of its non-local features can be systematically controlled.
This is carried out imposing that pairs of bonds for selection are
randomly chosen according to a probability that decays as a power-law
of their Manhattan distance, namely the distance measured as the
number of connections separating the sites in a regular lattice. This
physically plausible assumption is inspired on a geographical model
for complex networks where long-ranged shortcuts are incorporated to
regular lattices. Such a conceptual construction has been extensively
used as a way to explain the emergence of optimal navigation and
efficient transport in small-world systems~\cite{Navigation}. As a
consequence of the selection rule adopted here, we show that the
scaling properties of the system becomes dependent on the specific
value of the corresponding power-law exponent. A continuous variation
is then revealed from the traditional to the PR percolation behavior.
Moreover, the results of our extensive numerical simulations provide
strong evidence for the fact that the AER model, when applied to
regular lattices in two-dimensions, falls in the same universality
class as ordinary percolation.

Our bond percolation process takes place on a square lattice of size
$L$. At each step, two sites $i$ and $j$ are randomly selected with
probability $P(r_{ij})\sim r_{ij}^{-\alpha}$, where $r_{ij}$ is the
Manhattan distance between $i$ and $j$, measured as the number of
connections separating these sites in the underlying regular
lattice~\cite{footnote}. From each site $i$ and $j$, one bond is then
selected among its four adjacent edges, namely $e_{i}$ and $e_{j}$,
respectively. If at least one of these two bonds is already occupied,
the entire process of selection is restarted. If not, following the
product rule, weights are assigned to each of these bonds, in
proportionality to the product of the size (number of sites) of the
clusters they would potentially connect. In the case a bond connects
two sites in the same cluster, the weight is equal to the square of
the cluster size. The bond associated with the smallest weight becomes
occupied, while the other stays unoccupied, but can be selected again
in later steps.

Our model displays two distinct limiting behaviors, depending on the
exponent $\alpha$. For $\alpha=0$, we recover the usual PR, for which
the preliminary random selection of the bonds $e_{1}$ and $e_{2}$
constitutes a highly non--local process
\cite{Friedman2009,Grassberger2011}. In the limit of
$\alpha\rightarrow\infty$, the bonds $e_{1}$ and $e_{2}$ are always
adjacent, which corresponds to the AER process proposed in
Ref.~\cite{DSouza2010}, but applied here to regular lattices. Although
more spatially restricted than the PR process, the AER is still
non--local, since it requires information about the masses of the
joining clusters \cite{Grassberger2011}. As we show later, the finite
low-dimensionality of the square lattice employed here attenuates even
further the already weaker non--local features of the AER process.

\begin{figure}[!t]
  \includegraphics[width=7cm]{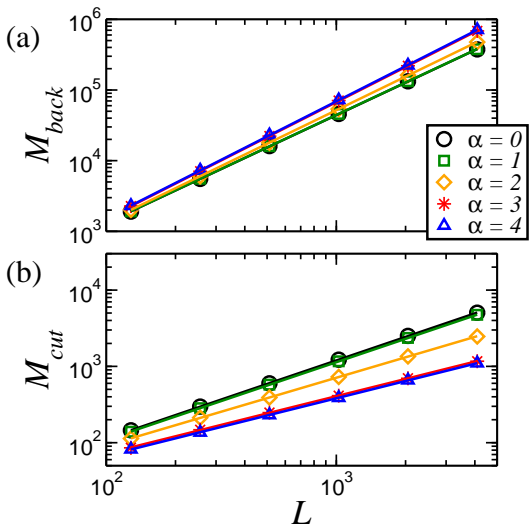}
  \caption{(Color online) (a) Log-log dependence of the mass of the
    conducting backbone $M_{back}$ on the system size for different
    values of the exponent $\alpha$. (b) the same as in (a), but for
    the number of cutting bonds $M_{cut}$. In both cases and for all
    values of $\alpha$, the evidence of scaling behavior substantiates
    the calculation of the fractal dimensions $d_{back}$ and $d_{cut}$
    as the slopes of the corresponding straight lines that are
    best--fitted to the simulation data. All quantities are averaged
    over at least $2500$ realizations precisely at the point in which
    the largest cluster appears.}
  \label{fig3}
\end{figure}

The percolation process stops when one among all clusters, namely the
spanning cluster, connects the lattice from top to bottom
\cite{Stauffer1992}. At that point, the fraction $p$ of occupied bonds
corresponds to the percolation threshold $p_{c}$. For $\alpha=0$, we
obtain $p_{c}=0.527\pm0.001$, which is in good agreement with previous
simulation results of the PR on the square lattice
\cite{Radicchi2010,Ziff2010}. Moreover, we find that $p_{c}$ decreases
smoothly and monotonically with $\alpha$ from this value to
$0.522\pm0.001$ at $\alpha=4$ (not shown). Next we apply the burning
algorithm \cite{Herrmann1984} to compute the mass of the spanning
cluster $M_{clus}$, the mass of its conducting backbone $M_{back}$,
and the mass (number) $M_{cut}$ of cutting bonds. The last ones, also
called red bonds, if removed, would break the spanning cluster in two,
therefore destroying the global connectivity of the system.

\begin{figure}[t]
  \includegraphics[width=7cm]{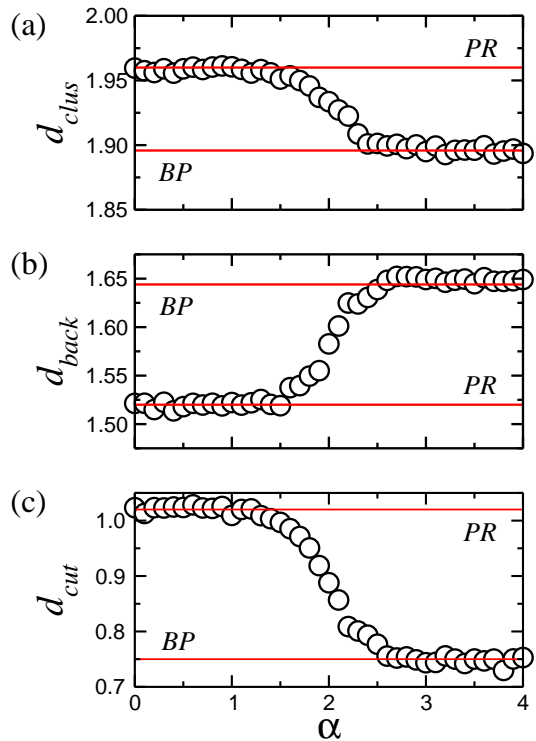}
  \caption{(Color online) Dependence on the exponent $\alpha$ of the
    size-scaling exponents for (a) the mass of the spanning cluster
    $d_{clus}$, (b) the mass of the conducting backbone $d_{back}$,
    and (c) the number of cutting bonds $d_{cut}$. In all cases, a
    crossover can be observed in the interval $1<\alpha<3$ from a
    regime of non-local explosive percolation at $\alpha=0$, to a
    regime that is compatible with ordinary bond percolation (BP), at
    sufficiently large values of $\alpha$. The dashed red lines
    correspond to $d_{clus}=1.96$ \cite{Radicchi2009,Ziff2009} and
    $91/48$ \cite{Stauffer1992} in (a), $d_{back}=1.52$
    \cite{Andrade2011} and $1.64$ \cite{Sahimi1994} in (b), and
    $d_{cut}=1.02$ \cite{Andrade2011} and $0.75$ \cite{Stauffer1992}
    in (c).}
  \label{fig4}
\end{figure}

Our results show that, regardless of the value of $\alpha$, all these
critical quantities scale with the system size $L$ as typical power
laws, $M_{back}\sim L^{d_{back}}$, $M_{cut}\sim L^{d_{cut}}$ (see
Fig.~\ref{fig3}), and $M_{clus}\sim L^{d_{clus}}$ (not shown), where
$d_{back}$, $d_{cut}$ and $d_{clus}$ are the fractal dimensions of the
conducting backbone, the cutting bonds, and the largest cluster,
respectively. In Figs.~\ref{fig4}(a)-(c) we show that all these
exponents exhibit a monotonic variation with $\alpha$, going from a
saturation regime of (non--local) explosive percolation at $\alpha=0$
to another compatible with ordinary bond percolation (BP) at
sufficiently large values of $\alpha$. Accordingly, for $\alpha=0$, we
recover the previously numerically calculated values of
$d_{clus}=1.96\pm0.01$ \cite{Radicchi2009,Ziff2009},
$d_{back}=1.52\pm0.03$, and $d_{cut}=1.02$ \cite{Andrade2011}. In all
three cases, by increasing $\alpha$, a crossover from PR to BP takes
place in the interval $1<\alpha<3$. More precisely, $d_{clus}$
decreases in this interval and starts fluctuating around $1.89$ for
$\alpha>3$ (see Fig.~\ref{fig4}a), in agreement with the classical
$2D$ value of $91/48$ \cite{Stauffer1992}. After increasing in the
interval $1<\alpha<3$, the exponents $d_{back}$ and $d_{cut}$ remain
practically constant for $\alpha>3$, around $1.64$ \cite{Sahimi1994}
(see Fig.~\ref{fig4}b) and $0.75$ \cite{Stauffer1992} (see
Fig.~\ref{fig4}c), respectively.  These values are fully compatible
with previously reported numerical calculations for ordinary (local)
percolation in $2D$. The variations of the exponents $d_{clus}$ and
$d_{back}$ within $1<\alpha<3$ reflect relevant changes in compactness
of the spanning cluster and its conducting backbone. Although the
spanning cluster becomes less compact as $\alpha$ increases
($d_{clus}$ decreases), the mass of the backbone $M_{back}$ tends to
occupy a larger fraction of $M_{clus}$, since the dimension $d_{back}$
increases in the same interval of $\alpha$ values, as shown in
Fig~\ref{fig3}. In these circumstances, a more compact conducting
backbone implies a smaller number of cutting bones (see
Figure~\ref{fig3}), therefore explaining the decrease in the exponent
$d_{cut}$.

\begin{table*}[ht]
  \begin{center}
\begin{tabular}{c|c|c|c|c|c}
$D$ & $p_{c,AP}$ & $p_{c,AER}$ & $d_{clus}$ - AP & $d_{clus}$ - AER & $d_{clus}$ - Classical~\cite{Stauffer1992} \\ \hline\hline
$2$ & $0.526550\pm0.000005$ & $0.52007\pm0.00001$ & $1.955\pm0.002$ & $1.899\pm0.001$ & $91/48$ \\
$3$ & $0.322096\pm0.000001$ & $0.285360\pm0.000008$ & $2.788\pm0.003$ & $2.530\pm0.003$ & $2.53$ \\
$4$ & $0.234160\pm0.000003$ & $0.202163\pm0.000004$ & $3.665\pm0.009$ & $3.079\pm0.005$ & $3.06$\\ 
$5$ & $0.184656\pm0.000006$ & $0.160454\pm0.000004$ & $4.61\pm0.01$ & $3.59\pm0.04$ & $3.54$\\
$6$ & $0.152642\pm0.000005$ & $0.134113\pm0.000002$ & $5.558\pm0.005$ & $4.46\pm0.01$ & $4$\\
\end{tabular}
\end{center}
\caption{Estimated values of the percolation threshold $p_{c}$ and the scaling 
  exponent $d_{clus}$ for hyper-cubic lattices of dimension $D$ calculated 
  using the jump method \cite{Manna2011,Jump}. The 
  presented values correspond to  averages over a minimum of 2500 realizations 
  of systems with sizes up to $L=4096$ ($D=2$), $256$ ($D=3$), $64$ ($D=4$), 
  $28$ ($D=5$), and $16$ ($D=6$).}
\label{table}
\end{table*}

Next we provide some analytical arguments that indicate how non--local
features are introduced in our percolation model through the
power--law probability $P(r_{ij})$. We first consider the average
distance between all pairs of sites in a empty lattice, $\langle
r\rangle=\sum_{r=1}^{L}r
N_{r}r^{-\alpha}/\sum_{r=1}^{L}N_{r}r^{-\alpha}$, where $N_{r}=4r$ is
the number of sites that are at a Manhattan distance $r$ from a given
site in the square lattice. Approximating the sum by an integral, we
obtain that $\langle r\rangle\sim\int_{1}^{L} r^{2-\alpha}dr$. It
follows that, for $\alpha<2$, $\langle r\rangle$ is limited by the
network size, leading to $\langle r\rangle\sim L$, while, for
$2\leq\alpha<3$, this distance scales as $\langle r\rangle\sim
L^{3-\alpha}$. For $\alpha\ge 3$ and sufficiently large lattice sizes,
$\langle r\rangle$ is always finite.  As a consequence, the effect of
non--locality on the scaling properties of the system would only play
a role for $\alpha<3$. In addition, distinct non-local behaviors
should be expected for the intervals $\alpha<2$ and $2\leq\alpha<3$.
These characteristics are consistent with the results displayed in
Fig.~\ref{fig4}. The observed mismatch between expected and
numerically calculated crossover values of the exponent $\alpha$ is a
consequence of finite-size scaling effects as well as the fact that
the sequential bond allocation leads to the presence of spatial
correlations in the percolation process. These correlations make the
assumption of an ever empty lattice, as adopted to compute $\langle
r\rangle$, no longer strictly valid.

In order to better confirm the role of non--locality on the PR
process, we perform additional simulations in the two limits of the
model at higher dimensions, namely for $\alpha=0$, which corresponds
to the original PR process, and for the AER process,
$\alpha\rightarrow\infty$. In these cases, improved performance can be
achieved by adopting the so-called jump method to analyze the behavior
of the order parameter $M_{clus}$ \cite{Manna2011,Jump}. For each
realization, we compute the average fraction $p$ of occupied bonds at
which a jump takes place, defined as the maximum change on $M_{clus}$
from the occupation of a single bond. This value of $p$ corresponds to
the percolation threshold $p_{c}$. The results for $p_{c}$ and
$d_{clus}$ in both limits and different dimensions are summarized in
Table~\ref{table}. Interestingly, the discrepancy between the fractal
dimensions calculated for PR and AER models increases substantially
with lattice dimensionality. Moreover, our calculations suggest that
the resemblance between regular BP and the limiting case
$\alpha\rightarrow\infty$ stands only up to five dimensions, when
compared with previous results from the literature
\cite{Stauffer1992,Sahimi1994}.

In summary, we have proposed a generalization of the PR model where
the range of non--locality in the percolation process can be
explicitly tuned. Our results show that this model displays a rich
variety of scaling behaviors, going from ordinary to non--local
explosive percolation. We expect our PR model, since it is based on a
geographical choice of bond pairs, to provide relevant physical
insights into the role of non--locality on the critical properties of
percolation.

We thank the Brazilian Agencies CNPq, CAPES, FUNCAP and FINEP, the
FUNCAP/CNPq Pronex grant, and the National Institute of Science and
Technology for Complex Systems in Brazil for financial support.


\begin{thebibliography}{00}
 
\bibitem{Stauffer1992}
  D. Stauffer and A. Aharony, {\it Introduction to Percolation Theory}
  (Taylor \& Francis, London, 1992).
%
\bibitem{Sahimi1994}
  M. Sahimi, {\it Applications of Percolation Theory}
  (Taylor \& Francis, London, 1994).
%
\bibitem{Perclass}
  M. Sahimi and S. Arbabi, Phys. Rev. Lett. {\bf 68},
  608 (1992); P. Grassberger and Y. C. Zhang, Physica A {\bf 224},
  169 (1999); J. S. Andrade {\it et al.}, Physica A {\bf 238},
  163 (1996); H. E. Stanley {\it et al.}, Physica A {\bf 266}
  5 (1999).
%
\bibitem{Percorr}
  S. Havlin {\it et al.}, Phys. Rev. Lett. {\bf 61}, 1438 (1988);
  S. Havlin {\it et al.}, Phys. Rev. A {\bf 40}. 1717 (1989);
  H. A. Makse {\it et al.}, Phys. Rev. E {\bf 53}, 5445 (1996);
  H. A. Makse {\it et al.}, Phys. Rev. E {\bf 54}, 3129 (1996).
\bibitem{Prakash1992}
  S. Prakash {\it et al.}, Phys. Rev. A {\bf 46}, R1724 (1992).
%
\bibitem{Achlioptas2009}
  D. Achlioptas, R. M. D'Souza, and J. Spencer, Science {\bf 323},
  1453 (2009).
%
\bibitem{Percexp}
  Y. S. Cho {\it et al.},
  Phys. Rev. Lett. {\bf 103}, 135702 (2009);
  Y. S. Cho, B. Kahng, and D. Kim,
  Phys. Rev. E {\bf 81}, R030103 (2010);
  Y. S. Cho {\it et al.},
  Phys Rev. E {\bf 82}, 042102 (2010);
  A. A. Moreira {\it et al.},
  Phys. Rev. E {\bf 81}, R040101 (2010);
  N. A. M. Ara\'ujo and H. J. Herrmann,
  Phys. Rev. Lett. {\bf 105}, 035701 (2010);
  N. A. M. Ara\'ujo {\it et al.},
  Phys. Rev. Lett. {\bf 106}, 095703 (2011);
  W. Chen and R. M. D'Souza,
  Phys. Rev. Lett. {\bf 106}, 115701 (2011);
  J. G\'omes-Gardenes {\it et al.},
  Phys. Rev. Lett. {\bf 106}, 128701 (2011);
  H. Hooyberghs and B. V. Schaeybroek,
  Phys. Rev. E {\bf 83}, 032101 (2011).

\bibitem{Friedman2009}
  E. J. Friedman and A. S. Landsberg, 
  Phys. Rev. Lett. {\bf 103}, 255701 (2009).
%
\bibitem{Ziff2009}
  R. M. Ziff,
  Phys. Rev. Lett. {\bf 103}, 045701 (2009).
%
\bibitem{Radicchi2009}
  F. Radicchi and S. Fortunato,
  Phys. Rev. Lett. {\bf 103}, 168701 (2009).
%
\bibitem{Radicchi2010}
  F. Radicchi and S. Fortunato,
  Phys. Rev. E {\bf 81}, 036110 (2010).
%
\bibitem{Ziff2010}
  R. M. Ziff,
  Phys. Rev. E {\bf 82},051105 (2010).
%
\bibitem{Manna2011}
  S. S. Manna and A. Chartterjee,
  Physica A {\bf 390}, 177 (2011).
%
\bibitem{Andrade2011}
  J. S. Andrade {\it et al.},
  Phys. Rev. E {\bf 83}, 031133 (2011).
%
\bibitem{Rozenfeld2010}
  H. D. Rozenfeld, L. K. Gallos, and H. A. Makse,
  Eur. Phys. J. B {\bf 75}, 305 (2010).
%
\bibitem{Kim2010}
  Y. Kim, Y. K. Yun and S. H. Yook,
  Phys. Rev. E {\bf 82}, 061105 (2010).

\bibitem{Grassberger2011}
  P. Grassberger {\it et al.},
  Phys. Rev. Lett. {\bf 106}, 225701 (2011);
%
\bibitem{Percont}
  R. A. da Costa {\it et al.},
  Phys. Rev. Lett {\bf 105}, 255701 (2010);
  H. K. Lee, B. J. Kim, and H. Park,
  Phys. Rev. E {\bf 84}, R020101 (2011);
  O. Riordan and L. Warnke,
  Science {\bf 333}, 322 (2011).
%
\bibitem{DSouza2010}
  R. M. D'Souza and M. Mitzenmacher,
  Phys. Rev. Lett. {\bf 104}, 195702 (2010).
\bibitem{Navigation}
  J. Kleinberg, Nature {\bf 406}, 845 (2000);
  M. Barthelemy, Phys. Rep. {\bf 499}, 1-101 (2011);
  G. Li {\it et al.}, Phys. Rev. Let. {\bf 104}, 018701 (2010);
  Y. Li {\it et al.}, Europhys. Lett. {\bf 92}, 58002 (2010);
  M. P. Viana and L. da F. Costa, Phys. Lett. A {\bf 375}, 1626-1629 (2011);
  L. K. Gallos, H. A. Makse, and M. Sigman, arXiv:1102.0604.
  
%
\bibitem{footnote} In fact, to account for the number $N_{r}$ of
  neighbors that are at a Manhattan distance $r$ from a particular
  site in a square lattice ($N_{r}=4r$), the random distance $r$ must
  be generated following a power-law distribution proportional to
  $r^{(1-\alpha)}$.
%
\bibitem{Herrmann1984}
  H. J. Herrmann, D. C. Hong, and H. E. Stanley,
  J. Phys. A {\bf17}, L261 (1984).
%
\bibitem{Jump}
  J. Nagler, A. Levina, and M. Timme,
  Nat. Phys. {\bf 7}, 265 (2011);
  K. J. Schrenk, N. A. M. Ara\'ujo, and H. J. Herrmann,
  Phys. Rev. E {\bf 84}, 041136 (2011).
%

\end{thebibliography}
\end{document}